\documentclass[onecolumn]{revtex4}
\usepackage{amsfonts}
\usepackage{amsmath}
\usepackage{amssymb}
\usepackage{charter}
\usepackage{graphicx}

\input{tcilatex}
\begin{document}

\title{M Times Photon Subtraction-Addition Coherent Superposition Operated
Odd-Schr\H{o}dinger-cat\ State: Nonclassicality and Decoherence }
\author{Li Huang, Qin Guo$^{\dag }\thanks{%
Corresponding author: guoqin91@163.com}$, Li-ying Jiang, Ge Chen, Xue-xiang
Xu, Wen Yuan}
\affiliation{$^{1)}$College of Physics and Communication Electronics, Jiangxi Normal
University, Nanchang, 330022, China}
\affiliation{$^{2)}$Key Laboratory of Optoelectronic and Telecommunication of Jiangxi,
Nanchang 330022, China \ }

\begin{abstract}
We introduce a new non-Gaussian state, generated by $m$ times coherent
superposition operation $a\cos \theta +a^{\dagger }e^{i\varphi }\sin \theta $%
\ (MCSO) on odd-Schr\H{o}dinger-cat\ state (OSCS). Its normalized constant
is turned out to be related with the Hermite polynomial. We further
investigate the nonclassical properties of the MCSO-OSCS through Mandel's
Q-parameter, quadrature squeezing, the photocount distribution and Wigner
function (WF). It is shown that the nonclassicality of the MCSO-OSCS is
influenced by the number of times ($m$) of coherent superpositon operation,
the angle $\theta $ and the amplitude $\left \vert \alpha _{0}\right \vert $%
. Especially the volume of negative region of WF increases with the
increment of parameters $m$, $\theta $ and $\alpha _{0}$. We also
investigate the decoherence of the MCSO-OSCS in terms of the fadeaway of the
negativity of WF in a thermal environment.

\textbf{Keywords: Non-Gaussian state. Wigner function. Coherent
Superposition Operation. odd-Schr\H{o}dinger-cat\ state. Decoherence. }

\textbf{PACS:} 0365, 0530, 4250
\end{abstract}

\maketitle

\section{Introduction}

Nonclassical states with non-Gaussian Wigner function (WF) have brought
great interest in quantum optics and quantum information science \cite{1}.
For example, a single-photon state with non-Gaussian behavior in phase space
has been found many applications in quantum information processing. In
particular, any single-mode nonclassical state has become a sufficient
resource to generate a two-mode entanglement via a beam-splitter \cite{1a}.
Recently, the non-Gaussian states have attracted more attention of both
experimentalists and theoreticians \cite{2,3,4,5,6}. It is possible to
generate and manipulate various non-Gaussian states through subtracting or
adding photon operation or photon subtraction-addition coherent
superposition operation on traditional quantum states or Gaussian states 
\cite{7}. For example, the photon subtraction transforms a Gaussian
entangled state (two-mode squeezed state) to a non-Gaussian entangled state
for a nonlocality test \cite{8} and entanglement distillation \cite{9}. The
photon addition can also transform a classical state to a nonclassical state 
\cite{10}. In laboratory the operation of photon subtraction or addition is
now realized practically \cite{11,12}. In Ref. \cite{13}, Lee \textit{et al.}
consider a coherent superposition of photon subtraction and addition, $%
ta+ra^{\dag }$, acting on a coherent state and a thermal state to form
non-Gaussian states, and propose the experimental scheme to implement this
elementary coherent operation. Furthermore, other non-Gaussian state is
obtained theoretically through $m$ times coherent superposition of photon
subtraction and addition, $\left( ta+ra^{\dag }\right) ^{m}$, acting on
thermal state \cite{13a} and coherent state \cite{13b}, respectively.

On the other hand, as a kind of nonclassical state, the so-called Schr\H{o}%
dinger cat states (SCS, quantum superpositions of coherent states \cite{14}%
), play an important role in fundamental tests of quantum theory \cite{15,16}
and in many quantum information processing tasks, including quantum
computation \cite{17}, quantum teleportation \cite{18} and precision
measurements \cite{19,20}. There have been a great deal of theoretical and
experimental attempts to generate a Schr\H{o}dinger-cat-type state and
considerable experimental progresses have been achieved in recent years \cite%
{21,22,23,24,25}. Such as in Ref. \cite{25} a Schr\H{o}dinger-cat-like state
is generated via a coherent superposition of photonic operations.

Thus an interesting question is naturally raised: can we operate the
coherent superposition operator $\left( a\cos \theta +a^{\dagger
}e^{i\varphi }\sin \theta \right) ^{m}$ on odd-Schr\H{o}dinger-cat state
(OSCS) to construct a new\ non-Gaussian quantum state? The answer is
definite. Considering the above reasons, we shall construct a new
nonclassical state (MCSO-OSCS) which is supposed to be realized in
experiment. In this paper, We focus on studying its nonclassical properties
of this state by deriving analytically some expressions, such as normalized
constant, sub-Poissonian statistics, photocount distribution and Wigner
function. In fact, systems are usually surrounded by a thermal reservoir,
and\ decoherence becomes an important topic in the fields of quantum optics.
Enlightened by these ideas, we shall also discuss its decoherence property
in a thermal environment in this paper.

The paper is organized as follows. In Sec. 2, the MCSO-OSCS is constructed
and its normalized constant turns out to be related with the Hermite
polynomial. In Sec. 3, the fidelity between MCSO-OSCS and its original state
(OSCS) shall be obtain. In Sec. 4, the nonclassical properties of the
MCSO-OSCS, such as sub-Poissonian statistics, quadrature squeezing
properties and photocount distribution are calculated analytically and then
discussed in details. In Sec.5, the explicitly analytical expression of WF
for the MCSO-OSCS is derived. According to the negativity of WF, the
nonclassical properties are also discussed in details. In Sec. 6, the
decoherence of the MCSO-OSCS in a thermal environment\ is investigated. In
Sec. 7, we end our work with main conclusions.

\section{Normalization of the MCSO-OSCS}

Theoretically, the MCSO-OSCS can be introduced by repeated application of
coherent superposition operator $\Omega $ to the OSCS ($\left \vert \alpha
_{0}\right \rangle -\left \vert -\alpha _{0}\right \rangle $) for $m$ times,
i.e.,%
\begin{equation}
\left \vert \psi _{m}\right \rangle =\Omega ^{m}\left( \left \vert \alpha
_{0}\right \rangle -\left \vert -\alpha _{0}\right \rangle \right) ,
\label{1}
\end{equation}%
\  \ where $\Omega =a\cos \theta +a^{\dagger }e^{i\varphi }\sin \theta $ with 
$\left[ a,a^{\dagger }\right] =1$ and $\theta \in \left( 0,\pi /2\right) $, $%
m$ is the order of coherent superposition operator (a non-negative integer), 
$\left \vert \alpha _{0}\right \rangle $ is a coherent state of amplitude $%
\left \vert \alpha _{0}\right \vert $. The density operator of the MCSO-OSCS
is $\rho _{m}=N_{m}^{-1}\left \vert \psi _{m}\right \rangle \left \langle
\psi _{m}\right \vert $, where $N_{m}$ is a normalized constant of the
MCSO-OSCS to be determined by $\mathtt{Tr}\rho _{m}=1$. If $\Omega ^{m}$
operates on the even SCS$\  \left( \left \vert \alpha _{0}\right \rangle
+\left \vert -\alpha _{0}\right \rangle \right) $, then we obtain the
MCSO-ESCS. We only discuss the properties of MCSO-OSCS in this paper, for an
odd SCS in general show stronger nonclassical properties than an even SCS 
\cite{25a}.

In order to obtain the normalized constant $N_{m}$, and note that the
operator $\Omega $ is not always Hermitian due to $\Omega \neq \Omega
^{\dagger }$ when $\cos \theta \neq e^{i\varphi }\sin \theta $, we shall
derive the normal ordering form of $\Omega ^{m}$ firstly. Recalling the
generating function of the Hermite polynomial $H_{m}(x)$ \cite{26}, i.e. $%
\sum \limits_{m=0}^{\infty }\frac{t^{m}}{m!}H_{m}(x)=\exp \left(
2xt-t^{2}\right) ,$ with 
\begin{equation}
H_{m}(x)=\sum \limits_{m=0}^{\left[ m/2\right] }\frac{\left( -1\right)
^{l}m!\left( 2x\right) ^{m-2l}}{l!\left( m-2l\right) !}=\frac{\partial ^{m}}{%
\partial t^{m}}\left. \exp \left( 2xt-t^{2}\right) \right \vert _{t=0},
\label{2}
\end{equation}%
using the Baker-Hausdorff formula $e^{A+B}=e^{A}e^{B}e^{-\frac{1}{2}%
[A,B]}=e^{B}e^{A}e^{-\frac{1}{2}[B,A]}$ \cite{27} and the technique of
integration within an ordered product (IWOP) of operators \cite{28}, we have%
\begin{eqnarray}
e^{\lambda \Omega } &=&\colon e^{\lambda \Omega +\frac{1}{2}\lambda
^{2}e^{i\varphi }\sin \theta \cos \theta }\colon  \notag \\
&=&\sum \limits_{m=0}^{\infty }\frac{\lambda ^{m}\left( -i\sqrt{\frac{1}{2}%
e^{i\varphi }\sin \theta \cos \theta }\right) ^{m}}{m!}\colon H_{m}(\frac{%
i\Omega }{\sqrt{2e^{i\varphi }\sin \theta \cos \theta }})\colon ,  \label{3}
\end{eqnarray}%
where the symbol $\colon \colon $stands for the normally ordering. Comparing
Eq.(\ref{3}) with the expansion of $e^{\lambda \Omega }$, i.e. $e^{\lambda
\Omega }=\sum \limits_{m=0}^{\infty }\frac{\lambda ^{m}}{m!}\Omega ^{m}$, we
can easily obtain the normal ordering form of $\Omega ^{m}$: 
\begin{equation}
\Omega ^{m}=\left( -i\sqrt{\frac{1}{2}e^{i\varphi }\sin \theta \cos \theta }%
\right) ^{m}\colon H_{m}(\frac{i\Omega }{\sqrt{2e^{i\varphi }\sin \theta
\cos \theta }})\colon .  \label{4}
\end{equation}

Similarly, $\Omega ^{\dagger m}=\left( a^{\dagger }\cos \theta
+ae^{-i\varphi }\sin \theta \right) ^{m}$ has the normal ordering form as
follows:

\begin{equation}
\Omega ^{\dagger m}=\left( -i\sqrt{\frac{1}{2}e^{-i\varphi }\sin \theta \cos
\theta }\right) ^{m}\colon H_{m}(\frac{i\Omega ^{\dagger }}{\sqrt{%
2e^{-i\varphi }\sin \theta \cos \theta }})\colon .  \label{5}
\end{equation}%
From Eq. (\ref{4}) and (\ref{5}), we also give the following relations 
\begin{equation}
\left \langle \beta \right \vert \Omega ^{m}\left \vert \alpha \right
\rangle =\left( -i\sqrt{\frac{1}{2}e^{i\varphi }\sin \theta \cos \theta }%
\right) ^{m}H_{m}(\frac{i\left( \alpha \cos \theta +\beta ^{\ast
}e^{i\varphi }\sin \theta \right) }{\sqrt{2e^{i\varphi }\sin \theta \cos
\theta }})\left \langle \beta \right \vert \left. \alpha \right \rangle ,
\label{6}
\end{equation}%
and 
\begin{equation}
\left \langle \beta \right \vert \Omega ^{\dagger m}\left \vert \alpha
\right \rangle =\left( -i\sqrt{\frac{1}{2}e^{-i\varphi }\sin \theta \cos
\theta }\right) ^{m}H_{m}(\frac{i\left( \beta ^{\ast }\cos \theta +\alpha
e^{-i\varphi }\sin \theta \right) }{\sqrt{2e^{-i\varphi }\sin \theta \cos
\theta }})\left \langle \beta \right \vert \left. \alpha \right \rangle ,
\label{7}
\end{equation}%
where $\left \vert \alpha \right \rangle $ and $\left \vert \beta
\right
\rangle $ are coherent states and $\left \langle \beta \right \vert
\left. \alpha \right \rangle =\exp \left[ -\frac{1}{2}\left( \left \vert
\alpha \right \vert ^{2}+\left \vert \beta \right \vert ^{2}\right) +\beta
^{\ast }\alpha \right] $ \cite{29,30}. Eq. (\ref{6}) and (\ref{7}) are very
useful in the following calculations.

Next, according to Tr$\rho _{m}=1$, we obtain%
\begin{equation}
N_{m}=\mathtt{Tr}\left[ \left \vert \psi _{m}\right \rangle \left \langle
\psi _{m}\right \vert \right] =\left \langle \psi _{m}\right \vert \left.
\psi _{m}\right \rangle .  \label{8}
\end{equation}%
Substituting Eq. (\ref{1}), (\ref{4})-(\ref{5}) into (\ref{8}), and
inserting the completeness\ relation of the coherent state $\int \frac{d^{2}z%
}{\pi }\left \vert z\right \rangle \left \langle z\right \vert =1,$
furthermore, with the help of Eq. (\ref{2}) and the following integral
formula

\begin{eqnarray}
&&\int \frac{d^{2}z}{\pi }\exp \left( \zeta \left \vert z\right \vert
^{2}+\xi z+\eta z^{\ast }+fz^{2}+gz^{\ast 2}\right)  \notag \\
&=&\frac{1}{\sqrt{\zeta ^{2}-4fg}}\exp \left( \frac{-\zeta \xi \eta +\xi
^{2}g+\eta ^{2}f}{\zeta ^{2}-4fg}\right) ,  \label{9}
\end{eqnarray}%
whose convergent condition is Re$(\zeta \pm f\pm g)<0$, Re$[(\zeta
^{2}-4fg)/(\zeta \pm f\pm g)]<0$, we obtain%
\begin{equation}
N_{m}=2\chi ^{m}\left[ \sum \limits_{k=0}^{m}\left( -1\right) ^{m}A\left
\vert H_{m-k}\left( B\right) \right \vert ^{2}-\sum \limits_{k=0}^{m}\left(
-1\right) ^{k}A\left \vert H_{m-k}\left( C\right) \right \vert ^{2}\exp
\left( -2\left \vert \alpha _{0}\right \vert ^{2}\right) \right] ,
\label{10}
\end{equation}%
\ where we have set 
\begin{eqnarray}
A &=&\left( 2\frac{\sin \theta }{\cos \theta }\right) ^{k}\frac{1}{k!}\left( 
\frac{m!}{\left( m-k\right) !}\right) ^{2},  \notag \\
B &=&i\sqrt{\frac{e^{-i\varphi }\sin \theta }{2\cos \theta }}\alpha _{0}+%
\frac{i\sqrt{e^{i\varphi }\cos \theta }}{\sqrt{2\sin \theta }}\alpha
_{0}^{\ast },  \notag \\
C &=&\frac{i\sqrt{e^{-i\varphi }\cos \theta }}{\sqrt{2\sin \theta }}\alpha
_{0}-\frac{i\sqrt{e^{i\varphi }\sin \theta }}{\sqrt{2\cos \theta }}\alpha
_{0}^{\ast },  \notag \\
\chi &=&-\frac{1}{2}\sin \theta \cos \theta ,  \label{11}
\end{eqnarray}%
and we have used the recurrence relation of $H_{m}\left( x\right) $:%
\begin{equation}
\frac{\partial }{\partial x^{l}}H_{m}\left( x\right) =\frac{2^{l}m!}{\left(
m-l\right) !}H_{m-l}\left( x\right) .  \label{12}
\end{equation}%
Eq.(\ref{10}) indicates that the normalization factor $N_{m}$ is just
related to a Hermite polynomial. Obviously, when $m=0$, the MCSO-OSCS just
reduces to the odd SCS. The analytical expression of $N_{m}$ is important
for further investigating the properties of MCSO-OSCS. For MCSO-ESCS, we can
change the negative sign "$-$" before the second sign of sum in Eq.((\ref{10}%
) to the positive sign "$+$" and obtain its normalized constant.

\section{Fidelity between MCSO-OSCS and OSCS}

In quantum teleportation, the fidelity $F$, which measures how close the
teleported state is to the original state, is the projection of the original
pure state $\left \vert \Psi _{in}\right \rangle $ of the density operator $%
\rho _{in}=\left \vert \Psi _{in}\right \rangle \left \langle \Psi
_{in}\right \vert $ onto the teleported state $\left \vert \Psi
_{out}\right
\rangle $ of the density operator $\rho _{out}:F=$ Tr$\left(
\rho _{out}\rho _{in}\right) $ \cite{31,32}. Here the fidelity measures how
close the new state (MCSO-OSCS) is to the original state (OSCS). The
fidelity between MCSO-OSCS (density matrix is $\rho _{m}$) and its original
OSCS ($\rho _{o}$) is defined as \cite{7} 
\begin{equation}
F_{\theta ,\varphi ,m}=\frac{\text{Tr}\left( \rho _{m}\rho _{o}\right) }{%
\text{Tr}\left( \rho _{o}^{2}\right) }.  \label{13}
\end{equation}%
In general, $0\leq F\leq 1$, and $F=1$ shows that the two states are same,
while $F=0$ shows that the two states are anamorphic absolutely. Employing
the similar procedure of deriving the normalization constant, the fidelity
for MCSO-OSCS can be calculated out as%
\begin{equation}
F_{\theta ,\varphi ,m}=\frac{N_{m}^{-1}\chi ^{m}}{4\left( 1-e^{-2\left \vert
\alpha _{0}\right \vert ^{2}}\right) ^{2}}\left \vert \left[ \left(
-1\right) ^{m}+1\right] \left[ H_{m}(B^{\ast })-e^{-2\left \vert \alpha
_{0}\right \vert ^{2}}H_{m}(C)\right] \right \vert ^{2}.  \label{14}
\end{equation}%
\begin{figure}[tbp]
\label{Fig1} \centering \includegraphics[width=0.8\columnwidth]{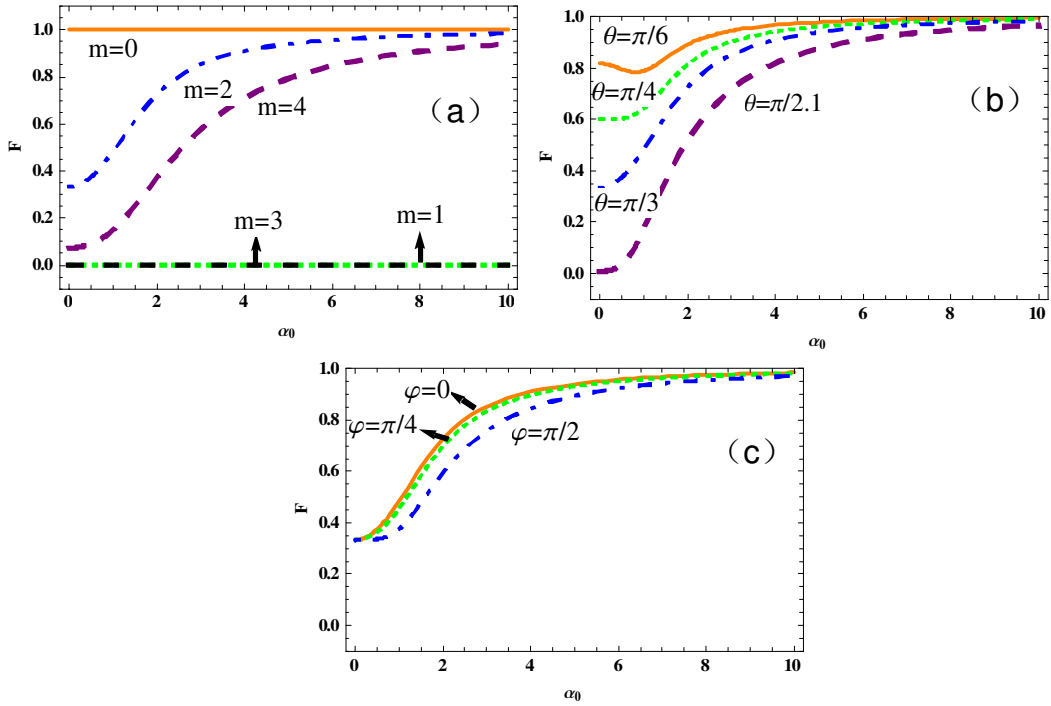}
\caption{Fidelity between MCSO-OSCS and OSCS as a function of $\protect%
\alpha _{0}$ (here $\protect \alpha _{0}$ is setted as real number) (a)\ $%
\protect \theta =\frac{\protect \pi }{3},$ $\protect \varphi =0,$ with
different $m$ values; (b) $m=2,$ $\protect \varphi =0,$ with different $%
\protect \theta $ values; (c) $m=2$, $\protect \theta =\frac{\protect \pi }{3},$
with different $\protect \varphi $ values. }
\end{figure}

In particular, when $m=0$, leading to $\rho _{m}=\rho _{o}$, then Eq. (\ref%
{13}) reduces to $F_{\theta ,\varphi ,m}=1$ (see Fig.1(a)), which indicates
that the MCSO-OSCS is reduced to the OSCS, as expected. In Eq. (\ref{14})
the fidelity $F_{\theta ,\varphi ,m}$ is equal to $0$ as $m$ is an odd
number because of the term $\left( \left( -1\right) ^{m}+1\right) $. In
Fig.1(a),\ we plot the fidelity $F_{\theta ,\varphi ,m}$ as the function of $%
\alpha _{0}$ for some different $m$ values with some given $\theta ,\varphi $
values, here $\alpha _{0}$ is setted as a real number, the same as in Figs.
2, 3, 8. It is obvious to note that $F_{\theta ,\varphi ,m}=0$ when $m=1,3$,
as expected, and $F_{\theta ,\varphi ,m}\neq 0$\ when $m=2,4$. When $m$ is
an even number, the fidelity increases monotonously with the increment of $%
\alpha _{0}$ and tends to $1$ finally, which indicates that the coherent
superposition operation has no influence on the filed when the field is
strong enough. Comparing with the curves of $m=0,2,4$, we find that the
smaller the value of $m$ is, the bigger the fidelity is. In order to see the
effect of different $\theta $ values on the fidelity, we plot the fidelity
as the function of $\alpha _{0}$ for some different $\theta $ values and
given $m,\varphi $\ values, see Fig.1(b). It is shown that the fidelity
decreases as $\theta $ increases. In addition, we study the relation of the
fidelity and parameter $\varphi $ through the plot of Fig.1(c). It is shown
that the values of parameter $\varphi $ have little influence on the
fidelity. It is also shown that the fidelity increases as the amplitude $%
\alpha _{0}$ increases from Fig.1(b) and 1(c).

\section{Nonclassical properties of MCSO-OSCS}

In this section, we shall discuss the nonclassical properties of the
MCSO-OSCS in terms of sub-Posissonian statistics, quadrature squeezing
properties and the negativity of its Wigner function.

\subsection{Mandel's Q-parameter}

The Mandel's Q-parameter measures the deviation of the variance of the
photon number distribution of the field state under consideration from the
Poissonian distribution of the coherent state, which has been defined as 
\cite{34}%
\begin{equation}
Q=\frac{\left \langle a^{\dagger 2}a^{2}\right \rangle }{\left \langle
a^{\dagger }a\right \rangle }-\left \langle a^{\dagger }a\right \rangle .
\label{15}
\end{equation}%
The quantum states has the Poissonian, sub-Poissonian and super-Poissonian
statistics for $Q=0,Q<0$ and $Q>0$, respectively. It is well known that the
negativity of $Q$-parameter refers to the nonclassical character of the
state, but a state may be nonclassical even though $Q$-parameter is positive
as pointed out in \cite{35}.

Using Eqs.(\ref{4}), (\ref{5}), $\rho _{m}=N_{m}^{-1}\left \vert \psi
_{m}\right \rangle \left \langle \psi _{m}\right \vert $ and IWOP technique
of operators, one can calculate $\left \langle a^{\dagger }a\right \rangle $
as

\begin{eqnarray}
\left \langle a^{\dagger }a\right \rangle &=&\text{Tr}\left( \rho
_{m}a^{\dagger }a\right)  \notag \\
&=&N_{m}^{-1}\chi ^{m}\sum \limits_{k=0}^{m}I\left[ 
\begin{array}{c}
+\left( -1\right) ^{m-k}\left( k+1+\left \vert \alpha _{0}\right \vert
^{2}\right) \left \vert H_{m-k}\left( B\right) \right \vert ^{2} \\ 
-\left( k+1-\left \vert \alpha _{0}\right \vert ^{2}\right) \left \vert
H_{m-k}\left( C\right) \right \vert ^{2}e^{-2\left \vert \alpha _{0}\right
\vert ^{2}} \\ 
+2\func{Re}\left[ R^{\ast }\alpha _{0}^{\ast }\left( m-k\right)
H_{m-k}\left( -B\right) H_{m-k-1}\left( B^{\ast }\right) \right] \\ 
-2\func{Re}\left[ R^{\ast }\alpha _{0}^{\ast }\left( m-k\right)
H_{m-k}\left( C^{\ast }\right) H_{m-k-1}\left( C\right) e^{-2\left \vert
\alpha _{0}\right \vert ^{2}}\right]%
\end{array}%
\right] -1,  \label{16}
\end{eqnarray}%
where%
\begin{eqnarray}
R &=&\frac{i\sqrt{2e^{-i\varphi }\sin \theta }}{\sqrt{\cos \theta }},  \notag
\\
I &=&\frac{2}{k!}\left( \frac{m!}{\left( m-k\right) !}\right) ^{2}\left(
-\left \vert R\right \vert ^{2}\right) ^{k},  \label{17}
\end{eqnarray}%
and get the value of $\left \langle a^{2}a^{\dagger 2}\right \rangle $ as%
\begin{equation}
\left \langle a^{2}a^{\dagger 2}\right \rangle =f_{1}\left( \alpha
_{0}\right) +f_{1}\left( -\alpha _{0}\right) -f_{2}\left( \alpha _{0}\right)
-f_{2}\left( -\alpha _{0}\right) ,  \label{18}
\end{equation}%
where 
\begin{eqnarray}
f_{1}\left( \alpha _{0}\right) &=&N_{m}^{-1}\chi ^{m}\left. \frac{\partial
^{2m+2}}{\partial t^{m}\partial s^{m}\partial \lambda \partial \eta }\frac{%
e^{-\left \vert \alpha _{0}\right \vert ^{2}-K^{\ast }t+Ks-s^{2}-t^{2}}}{%
\sqrt{1-4\lambda \eta }}e^{\left( H+H_{0}+\left \vert \alpha _{0}\right
\vert ^{2}\right) /\left( 1-4\lambda \eta \right) }\right \vert
_{s=t=\lambda =\eta =0},  \notag \\
f_{2}\left( \alpha _{0}\right) &=&N_{m}^{-1}\chi ^{m}\left. \frac{\partial
^{2m+2}}{\partial t^{m}\partial s^{m}\partial \lambda \partial \eta }\frac{%
e^{-\left \vert \alpha _{0}\right \vert ^{2}-K^{\ast }t-Ks-s^{2}-t^{2}}}{%
\sqrt{1-4\lambda \eta }}e^{\left( H-L_{0}-\left \vert \alpha _{0}\right
\vert ^{2}\right) /\left( 1-4\lambda \eta \right) }\right \vert
_{s=t=\lambda =\eta =0},  \label{19}
\end{eqnarray}%
and%
\begin{eqnarray}
K &=&i\frac{\sqrt{2e^{-i\varphi }\cos \theta }}{\sqrt{\sin \theta }}\alpha
_{0},  \notag \\
H &=&-\left \vert R\right \vert ^{2}ts+R^{2}t^{2}\eta +\alpha _{0}^{\ast
2}\eta +\alpha _{0}^{2}\lambda +R^{\ast 2}s^{2}\lambda ,  \notag \\
H_{0} &=&R\alpha _{0}t-R^{\ast }\alpha _{0}^{\ast }s+2R\alpha _{0}^{\ast
}t\eta -2R^{\ast }\alpha _{0}\lambda s,  \notag \\
L_{0} &=&R\alpha _{0}t+R^{\ast }\alpha _{0}^{\ast }s-2R\alpha _{0}^{\ast
}t\eta -2R^{\ast }\alpha _{0}\lambda s.  \label{20}
\end{eqnarray}%
Here $s,t,\lambda ,\eta $ are parameters introduced into the calculation
process and will be eliminated after finishing the calculation by setting
them to zero. Furthermore, one can use the relation $\left[ a,a^{\dagger }%
\right] =1$ to obtain 
\begin{equation}
\left \langle a^{\dagger 2}a^{2}\right \rangle =\left \langle
a^{2}a^{\dagger 2}\right \rangle -4\left \langle a^{\dagger }a\right \rangle
-2.  \label{21}
\end{equation}%
\begin{figure}[tbp]
\label{Fig2} \centering \includegraphics[width=0.8\columnwidth]{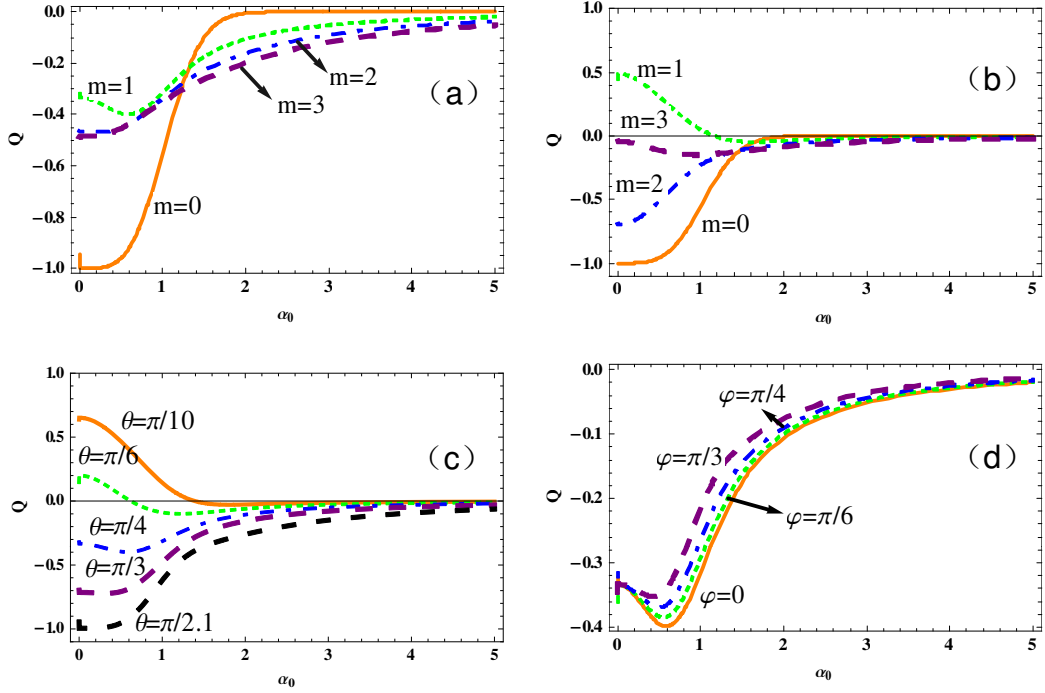}
\caption{Mandel's Q-parameter of MCSO-OSCS as a function of $\protect \alpha %
_{0}$ (a)\ $\protect \theta =\frac{\protect \pi }{4}$, $\protect \varphi =0,$
with different $m$ values; (b) $\protect \theta =\frac{\protect \pi }{8},%
\protect \varphi =0,$ with different $m$ values; (c) $m=1,$ $\protect \varphi %
=0,$with different $\protect \theta $ values; (d) $m=1,$ $\protect \theta =%
\frac{\protect \pi }{4},$with $\protect \varphi =\frac{\protect \pi }{3},\frac{%
\protect \pi }{4},\frac{\protect \pi }{6},0$ (from upper to lower curves). }
\end{figure}

Substituting Eqs. (\ref{16}), and (\ref{21}) into (\ref{15}), and using the
method of numerical calculation we can study the property of Mandel's
Q-parameter for the MCSO-OSCS. The Q-parameters of MCSO-OSCS as the function
of $\alpha _{0}$ are depicted in Fig. 2 for several different values of $m$, 
$\theta $ and $\varphi $. It is interesting to note that the values of
Q-parameter are always smaller than zero for different $m$ values under the
given $\theta =\frac{\pi }{4}$ and $\varphi =0$ (see Fig. 2(a)), which
indicates sub-Poissonian statistics. In addition, the absolute value of
Q-parameter decreases with the increment of $\alpha _{0}$ till tends to
zero, which indicates that all states under different $m$ values will tend
to the Poissonian statistics (the distribution of a coherent state) when the
value of $\alpha _{0}$ is big enough. However, we can see that the range of
the Q-parameter is $\left[ -1,0.5\right] $ in Fig. 2(b)\ with different $m$
values and for given $\theta =\frac{\pi }{8}$ and $\varphi =0$. That
indicates the\ MCSO-OSCS with small value of $\theta $ may do not exhibit
sub-Poissonian statistics but exhibit super-Poissonian statistics.

In particular, when $m=1$ and $\varphi =0$, the MCSO-OSCS deduces to the
COSCS ($\Omega \left( \left \vert \alpha _{0}\right \rangle -\left \vert
-\alpha _{0}\right \rangle \right) $) \cite{25a}. From Fig. 2(c), We can see
that the range of the Q-parameter is $\left[ -1,0.7\right] $. From the
criteria of Q-parameter, one finds that the MCSO-OSCS exhibits the
sub-Poissonian statistics for $\theta =\frac{\pi }{4}$, $\frac{\pi }{3}$ and 
$\frac{\pi }{2.1}$. In the area of $\theta \in (0,\frac{\pi }{2})$, the
bigger the value of $\theta $ is, the more chance the state exhibits the
sub-Poissonian statistics. The similar conclusion can also be seen in Ref. 
\cite{25a}. In addition, from Fig. 2(d) we can see that the absolute value
of Q-parameter decreases as $\varphi $\ increases, but this difference is
not obvious.

\subsection{Quadrature squeezing properties of MCSO-OSCS}

One observes nonclassical effects not only through sub-Poissonian statistics
but also through squeezing effects, which do not allow classical
interpretation of photoelectric counting events. Here, we consider an
appropriate quadrature operator $X_{\theta }=ae^{-i\theta }+a^{\dagger
}e^{i\theta }$, and the squeezing can be characterized by $\left \langle
\left( \Delta X_{\theta }\right) ^{2}\right \rangle _{\min }<1$ with respect
to angle $\theta $, or by the normal ordering form $\left \langle \colon
\left( \Delta X_{\theta }\right) ^{2}\colon \right \rangle _{\min }<0$ \cite%
{39}. Upon expanding the terms in $\left \langle \colon \left( \Delta
X_{\theta }\right) ^{2}\colon \right \rangle _{\min }$, one can minimize its
value over the whole angle $\theta $, which is given by \cite{40} 
\begin{equation}
S=\left \langle \colon \left( \Delta X_{\theta }\right) ^{2}\colon \right
\rangle _{\min }=-2\left \vert \left \langle a^{\dagger 2}\right \rangle
-\left \langle a^{\dagger }\right \rangle ^{2}\right \vert +2\left \langle
a^{\dagger }a\right \rangle -2\left \vert \left \langle a^{\dagger }\right
\rangle \right \vert ^{2}.  \label{23}
\end{equation}%
Then its negative value in the range $\left[ -1,0\right) $ indicates
squeezing (or nonclassicality). Similarly, using the integration formula (%
\ref{9}), we obtain 
\begin{equation}
\left \langle a^{\dagger }\right \rangle =0,  \label{24}
\end{equation}%
and%
\begin{equation}
\left \langle a^{\dagger 2}\right \rangle =f_{3}\left( \alpha _{0}\right)
+f_{3}\left( -\alpha _{0}\right) -f_{4}\left( \alpha _{0}\right)
-f_{4}\left( -\alpha _{0}\right) ,  \label{25}
\end{equation}%
where%
\begin{eqnarray}
f_{3}\left( \alpha _{0}\right) &=&N_{m}^{-1}\chi ^{m}\frac{\partial ^{2m}}{%
\partial t^{m}\partial s^{m}}\left. \left( Rt+\alpha _{0}^{\ast }\right)
^{2}e^{-t^{2}-s^{2}-\left \vert R\right \vert ^{2}st+2Bt-2B^{\ast }s}\right
\vert _{s=t=0},  \notag \\
f_{4}\left( \alpha _{0}\right) &=&N_{m}^{-1}\chi ^{m}\frac{\partial ^{2m}}{%
\partial t^{m}\partial s^{m}}\left. \left( Rt+\alpha _{0}^{\ast }\right)
^{2}e^{-t^{2}-s^{2}-\left \vert R\right \vert ^{2}st-2C^{\ast }t-2Cs-2\left
\vert \alpha _{0}\right \vert ^{2}}\right \vert _{s=t=0}.  \label{26}
\end{eqnarray}

Using Eqs.(\ref{16}), (\ref{24}) , (\ref{25}) and (\ref{23}), one can obtain
the expression of the quadrature squeezing $S$ of MCSO-OSCS. 
\begin{figure}[tbp]
\label{Fig3} \centering \includegraphics[width=0.8\columnwidth]{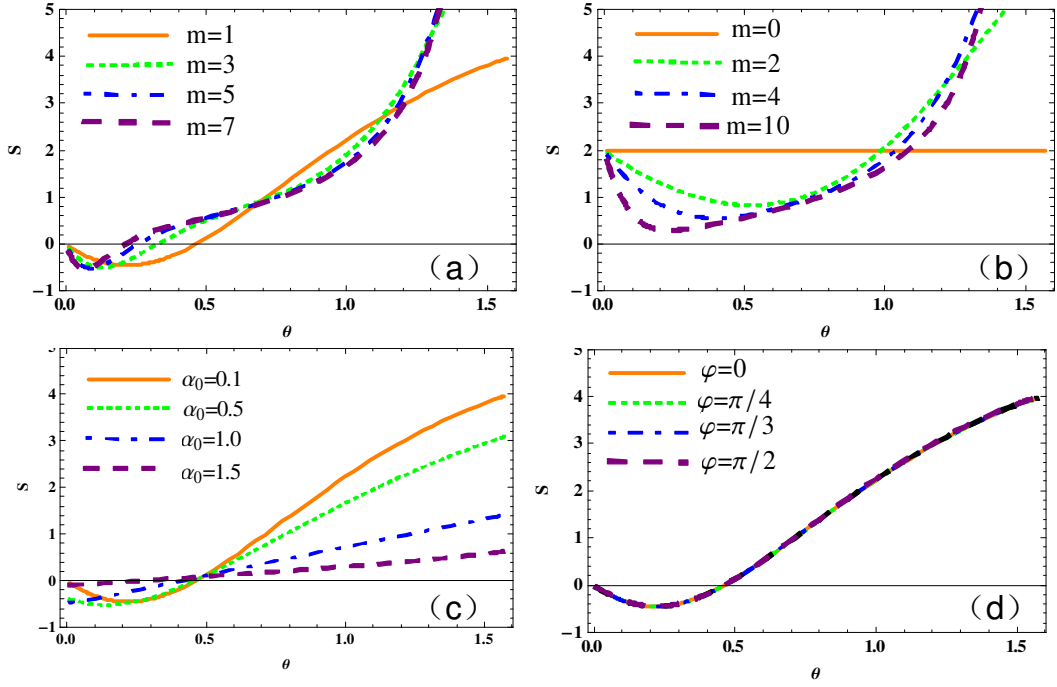}
\caption{Quadrature squeezing of MCSO-OSCS as a function of $\protect \theta $
(a)\ $\protect \alpha _{0}=0.1$, $\protect \varphi =0$, with different $m$
values; (b) $\protect \alpha _{0}=0.1$, $\protect \varphi =0$, with different $%
m$ values; (c) $m=1$, $\protect \varphi =0,$with different $\protect \alpha %
_{0}$ values; (d) $m=1$, $\protect \alpha _{0}=0.1$, with different $\protect%
\varphi $ values. }
\end{figure}

We plot the graph of quadrature squeezing $S$ as a function of $\theta $ for
some different $m$ values and for given $\varphi $ and $\alpha _{0}$ values,
(say $\varphi =0$ and $\alpha _{0}=0.1$),\ see Fig. 3(a) and (b). It is
interesting to find that the MCSO-OSCS can exhibit squeezing when the
parameter $m$ is odd ($m=1,3,5,7$) and the angle $\theta $ is smaller than a
threshold, while can't exhibit squeezing when the parameter $m$ is even ($%
m=0,2,4,10$) for any angle $\theta $. Furthermore, we find that the original
state $\left( m=0\right) $ can't exhibit squeezing, which implies that the
odd times coherent superposition operation ($\Omega ^{m},m$ is odd.) can
achieve squeezing. Small angle $\theta $ corresponds to the case that the
subtracting photon operation is in the ascendant, which indicates that
subtracting photon operation is benefit to squeezing under the case of odd $%
m $.

In Fig. 3(c), we plot the graph of $S$ as a function of $\theta $ for some
different $\alpha _{0}$ values and for given $\varphi $ and $m$ values, (say 
$\varphi =0$ and $m=1$). We find that small value of $\alpha _{0}$ is
helpful to squeezing on condition that the angle $\theta $ is smaller than a
threshold. From Fig. 3(d), We can see that different$\  \varphi $ values have
no effect on the squeezing of MCSO-OSCS.

\subsection{Photocount Distribution of MCSO-OSCS}

For the case of a single radiation mode of registering $n$ photoelectrons in
the time interval $T$, the photon counting distribution $P\left( n\right) $
is given by \cite{41}, 
\begin{equation}
P\left( n\right) =Tr\left[ \rho \colon \frac{\left( \xi a^{\dagger }a\right)
^{n}}{n!}e^{-\xi a^{\dagger }a}\colon \right] ,  \label{27}
\end{equation}%
where $\xi \propto T$ is called the quantum efficiency (a measure) of the
detector, $\rho $ is a single-mode density operator of the light field
concerned. When $\xi =1$, $P\left( n\right) $ becomes the photon number
distribution (PND)\ for a given state. By virtue of the technique of IWOP of
operators, Fan and Hu deduce a reformed formula as showed in reference \cite%
{42}, 
\begin{equation}
P\left( n\right) =\frac{\xi ^{n}}{\left( \xi -1\right) ^{n}}\int \frac{d^{2}z%
}{\pi }e^{-\xi \left \vert z\right \vert ^{2}}L_{n}\left( \left \vert
z\right \vert ^{2}\right) Q\left( \sqrt{1-\xi }z\right) ,  \label{28}
\end{equation}%
where $Q\left( \beta \right) =\left \langle \beta \right \vert \rho
\left
\vert \beta \right \rangle $ is the Q-function, $\left \vert \beta
\right
\rangle $ is the coherent state, and $L_{n}\left( x\right) $ is the
Laguerre polynomials. Once the Q-function of $\rho $ is known, it is easy to
calculate the photocount distribution of MCSO-OSCS from Eq.(\ref{28}).

The Q-function of MCSO-OSCS is given by 
\begin{eqnarray}
Q\left( \beta \right) &=&\left \langle \beta \right \vert \rho _{m}\left
\vert \beta \right \rangle  \notag \\
&=&N_{m}^{-1}\chi m\left( \left \langle \beta \right \vert \colon
H_{m}(a,a^{\dag })\colon \left( \left \vert \alpha _{0}\right \rangle -\left
\vert -\alpha _{0}\right \rangle \right) \left( \left \langle \alpha
_{0}\right \vert -\left \langle -\alpha _{0}\right \vert \right) \colon
H_{m}^{\ast }(a,a^{\dag })\colon \left \vert \beta \right \rangle \right) ,
\label{29}
\end{eqnarray}%
where $H_{m}(a,a^{\dag })=H_{m}(i\Omega /\sqrt{2e^{i\varphi }\sin \theta
\cos \theta })$. Then substituting Eq.(\ref{29}) into Eq.(\ref{28}) and
using Eq.(\ref{9}) and the two-variable Hermite polynomials expression of
Laguerre polynomials \cite{26}%
\begin{equation}
L_{n}\left( zz^{\ast }\right) =\frac{\left( -1\right) ^{n}}{n!}H_{n,n}\left(
z,z^{\ast }\right) =\frac{\left( -1\right) ^{n}}{n!}\left. \frac{\partial
^{2n}}{\partial \mu ^{n}\partial \nu ^{n}}e^{-\mu \nu +\mu z+\nu z^{\ast
}}\right \vert _{\mu =\nu =0},  \label{30}
\end{equation}%
we obtain the final result of $P\left( n\right) $ 
\begin{equation}
P\left( n\right) =2T_{m,n}\sum \limits_{j=0}^{m}\sum
\limits_{l,k=0}^{n}A_{j,k} \left[ 
\begin{array}{c}
e^{-\xi \left \vert \alpha _{0}\right \vert ^{2}}H_{m-l-j}\left( \frac{%
K-J^{\ast }}{2}\right) H_{m-j-k}\left( \frac{J-K^{\ast }}{2}\right) \\ 
-\left( -1\right) ^{n-k}e^{\left( \xi -2\right) \left \vert \alpha
_{0}\right \vert ^{2}}H_{m-l-j}\left( \frac{K^{\ast }+J}{2}\right)
H_{m-j-k}\left( \frac{K+J^{\ast }}{2}\right)%
\end{array}%
\right] ,  \label{31}
\end{equation}%
where we have set 
\begin{eqnarray}
A_{j,k} &=&\frac{\left( -1\right) ^{j+k}G^{j+l}G^{\ast j+k}F^{n-l}F^{\ast
n-k}}{l!j!k!\left( m-l-j\right) !\left( n-l\right) !\left( n-k\right)
!\left( m-j-k\right) !},  \notag \\
F &=&\sqrt{1-\xi }\alpha _{0},G=-\sqrt{1-\xi }R^{\ast },  \notag \\
J &=&\left( 1-\xi \right) R\alpha _{0},T_{m,n}=N_{m}^{-1}\chi ^{m}\frac{%
n!\left( m!\right) ^{2}\xi ^{n}}{\left( 1-\xi \right) ^{n}}.  \label{32}
\end{eqnarray}%
\begin{figure}[tbp]
\label{Fig4} \centering \includegraphics[width=0.7\columnwidth]{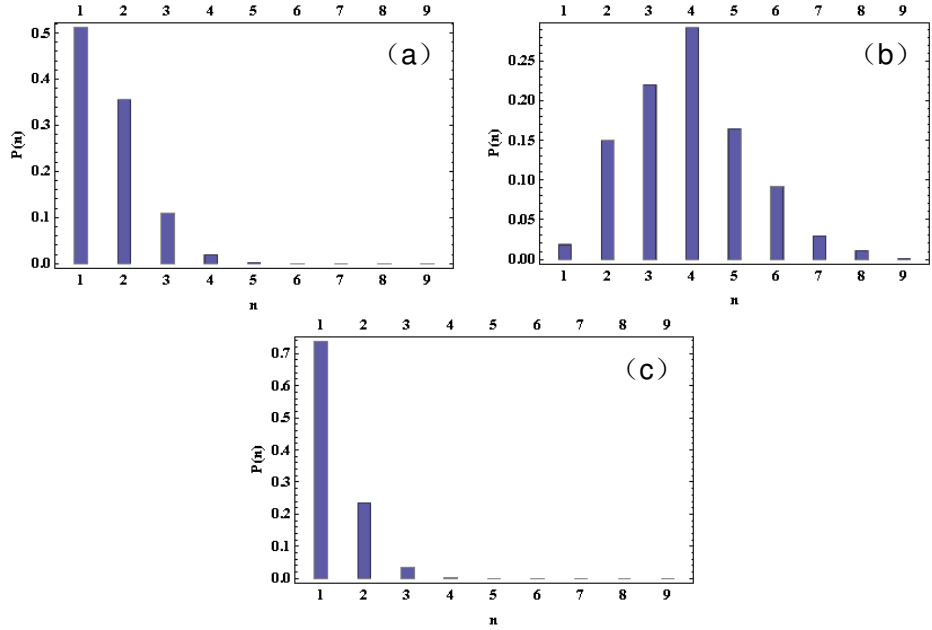}
\caption{Photocount distribution of MCSO-OSCS as a function of $n$ (a
non-negative integer) for $\protect \alpha _{0}=0.5+i0.5$ (a) $m=4$, $\protect%
\theta =\frac{\protect \pi }{4}$, $\protect \varphi =0$, $\protect \xi =0.2$;
(b) $m=4$, $\protect \theta =\frac{\protect \pi }{4}$, $\protect \varphi =0$, $%
\protect \xi =0.9$; (c) $m=1$, $\protect \theta =\frac{\protect \pi }{4}$, $%
\protect \varphi =0$, $\protect \xi =0.2$.}
\end{figure}
\begin{figure}[tbp]
\label{Fig5} \centering \includegraphics[width=0.7\columnwidth]{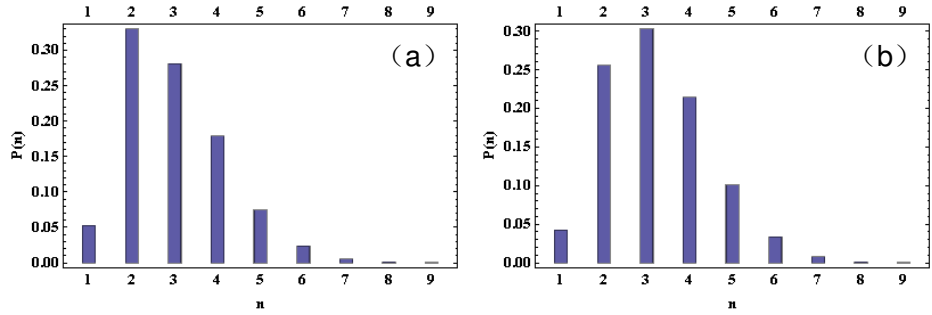}
\caption{Photocount distribution of MCSO-OSCS as a function of $n$ (a
non-negative integer) for $\protect \alpha _{0}=0.5+i0.5$ (a) $m=4$, $\protect%
\theta =\frac{\protect \pi }{8}$, $\protect \varphi =0$, $\protect \xi =0.9$;
(b)$m=4$, $\protect \theta =\frac{\protect \pi }{8}$, $\protect \varphi =\frac{%
\protect \pi }{2}$, $\protect \xi =0.9$.}
\end{figure}

In order to discuss the photocount distribution of MCSO-OSCS, we plot the
graph of $P\left( n\right) $ for several given parameters $\varphi ,$ $%
\theta $, $\alpha _{0}$, $m$, or $\xi $ in Figs. 4 and 5. Comparing with
Figs. 4(a) and (b), we find that for some given values of $m,$ $\alpha _{0},$
$\varphi ,$ and $\theta $, the corresponding probability-peak of photocount
distribution moves from $n=1$ to $n=4$ as $\xi =0.2$ increases to $0.9$,
which means that the probability of registering big photon-numbers is
increasing gradually while the probability of registering small
photon-numbers is decreasing when we increase the time interval $T$.
Meanwhile, the larger the $\xi $ is, the wider tail of photocount
distribution of MCSO-OSCS has. We can also see that the probability of
finding big photon-numbers increases with the increment of the parameter $m$
(see Figs. 4(a) and 4(c)) or $\theta $ (see Figs. 4(b) and 5(a)). Similarly,
the probability of finding big photon-numbers increases with the increment
of parameter $\varphi $ (see Figs. 5(a) and 5(b)).

\section{Wigner Function of the MCSO-OSCS}

The WF is a quasi-probability distribution, which fully describes the state
of a quantum system in phase space. The partial negativity of the WF is
indeed a good indication of the highly nonclassical character of the state 
\cite{43}. Therefore it is worth obtaining the WF for any states and using
the negative region to check whether a state has nonclassicality. For a
single-mode system, the WF $W\left( \alpha ,\alpha ^{\ast }\right) $
associated with a quantum state density matrix $\rho $ can be expressed as 
\cite{44}:

\begin{equation}
W\left( \alpha \right) =\frac{1}{\pi }e^{2\left \vert \alpha \right \vert
^{2}}\int \frac{d^{2}z}{\pi }\left \langle -z\right \vert \rho \left \vert
z\right \rangle e^{-2\left( \alpha ^{\ast }z-\alpha z^{\ast }\right) },
\label{33}
\end{equation}%
where $\left \vert z\right \rangle $ is the coherent state. Substituting $%
\rho _{m}=N_{m}^{-1}\left \vert \psi _{m}\right \rangle \left \langle \psi
_{m}\right \vert $ into Eq.(\ref{33}), we can finally obtain the WF of
MCSO-OSCS:

\begin{equation}
W\left( \alpha \right) =W_{\alpha _{0}}\left( \alpha \right) +W_{-\alpha
_{0}}\left( \alpha \right) -W_{\alpha _{0}}^{\prime }\left( \alpha \right)
-W_{-\alpha _{0}}^{\prime }\left( \alpha \right) ,  \label{34}
\end{equation}%
where we have set 
\begin{eqnarray}
W_{\alpha _{0}}\left( \alpha \right) &=&\sum \limits_{k=0}^{m}D_{m}\left(
-1\right) ^{k}e^{-2\left \vert \alpha -\alpha _{0}\right \vert ^{2}}\left
\vert H_{m-k}\left( -C^{\ast }+R\alpha \right) \right \vert ^{2},  \notag \\
W_{\alpha _{0}}^{\prime }\left( \alpha \right) &=&\sum
\limits_{k=0}^{m}D_{m}\left( -1\right) ^{m}e^{2\alpha _{0}^{\ast }\alpha
-2\alpha ^{\ast }\alpha _{0}-2\left \vert \alpha \right \vert
^{2}}H_{m-k}\left( B^{\ast }-R^{\ast }\alpha ^{\ast }\right) H_{m-k}\left(
B+R\alpha \right) ,  \label{35}
\end{eqnarray}%
and 
\begin{equation}
D_{m}=\frac{1}{\pi }N_{m}^{-1}\left( \frac{1}{2}\sin \theta \cos \theta
\right) ^{m}\left( 2\frac{\sin \theta }{\cos \theta }\right) ^{k}\frac{1}{k!}%
\left( \frac{m!}{\left( m-k\right) !}\right) ^{2}.  \label{36}
\end{equation}

It is found that the sum of $W_{\alpha _{0}}^{\prime }\left( \alpha \right) $
and $W_{-\alpha _{0}}^{\prime }\left( \alpha \right) $ is a real function
due to $W_{-\alpha _{0}}^{\prime }\left( \alpha \right) =\left[ W_{\alpha
_{0}}^{\prime }\left( \alpha \right) \right] ^{\ast }$. By using Eq. (\ref%
{34}), the WFs as a function of real and imaginary parts of $\alpha $ for
several different values of $m$, $\alpha _{0}$ and $\theta $ are depicted in
Figs. 6 and 7. 
\begin{figure}[tbp]
\label{Fig6} \centering \includegraphics[width=0.8\columnwidth]{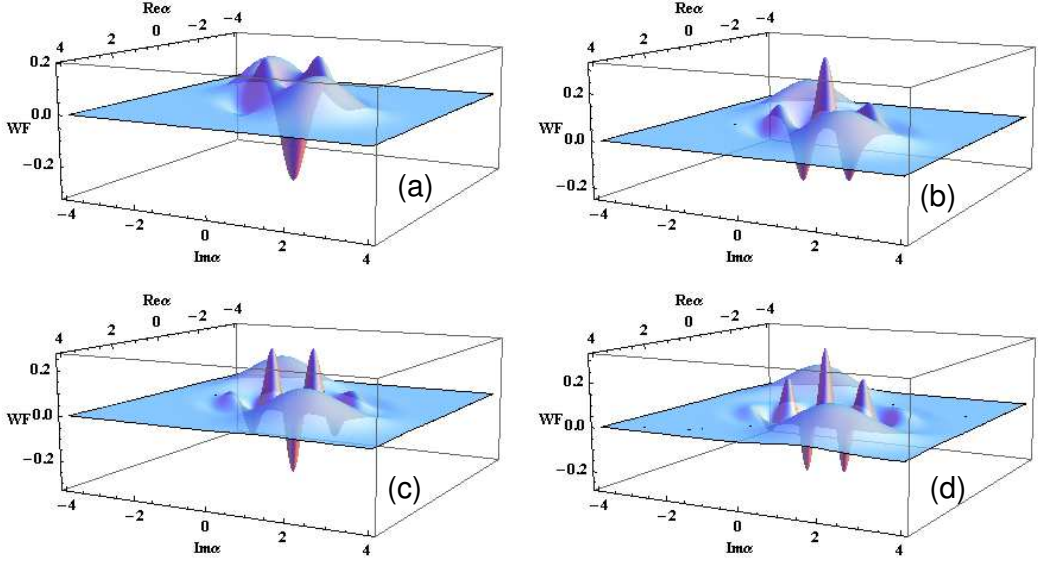}
\caption{Wigner function distributions of MCSO-OSCS with $\protect \varphi =0,%
\protect \theta =\frac{\protect \pi }{3},\protect \alpha _{0}=1+i$ (a) $m=0$;
(b) $m=1$; (c) $m=2$; (d) $m=3$.}
\end{figure}
\begin{figure}[tbp]
\label{Fig7} \centering \includegraphics[width=0.8\columnwidth]{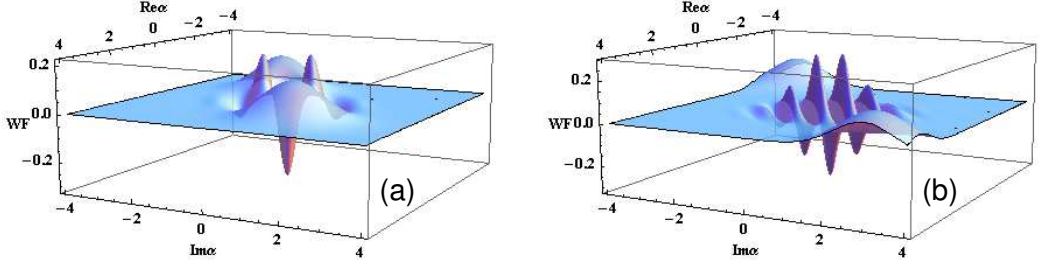}
\caption{Wigner function distributions of MCSO-OSCS with $m=2,\protect%
\varphi =0$ (a) $\protect \alpha _{0}=1+i,\protect \theta =\frac{\protect \pi }{%
8}$; (b) $\protect \alpha _{0}=2(1+i),\protect \theta =\frac{\protect \pi }{3}$%
. }
\end{figure}

We can see clearly that the figures of WF distribution are non-Gaussian. In
addition, as evidence of the nonclassicality of the state, it is easy to see
that there is a negative region of the WF in each plot. From Fig. 6, We can
see\ that the figures of WF exist odd (even) negative peaks when the values
of $m$ are even (odd) for given $\alpha _{0}$, $\varphi $ and $\theta $, and
exhibit more vibration character as the value of $m$ increasing. Meanwhile,
we can find that the minimum value of the WF occurs at the center of the
figure when $m$ is an even number (see Fig. 6(a) and Fig. 6(c)). But the
case is not true when $m$ is an odd number (see Fig. 6(b) and Fig. 6(d)).
Comparing Fig. 6(c) ($\theta =\frac{\pi }{3}$, $m=2$, $\alpha _{0}=1+i$)
with Fig. 7(a) ($\theta =\frac{\pi }{8}$, $m=2$, $\alpha _{0}=1+i$), we can
see that the width of\ the figure of WF in one direction increases as
increasing the value of $\theta $. Comparing Fig. 6(c) ($\alpha _{0}=1+i$, $%
m=2$, $\theta =\frac{\pi }{3}$) with Fig. 7(b) ($\alpha _{0}=2+2i,m=2,\theta
=\frac{\pi }{3}$), we can also see that the figure of WF also\ exhibits more
vibration character as increasing the value of amplitude $\left \vert \alpha
_{0}\right \vert $.

The volume of the negative part of the WF were used in \cite{45,46} to
describe the interference effects which determine the departure from
classical behavior. In order to further evaluate how these parameters $m$, $%
\alpha _{0}$, and $\theta $ affect the negative part of WF distribution for
MCSO-OSCS, we shall consider the negative part volume of WF which may be
written as%
\begin{equation}
\delta =\frac{1}{2}\left[ \int d^{2}\alpha \left \vert W\left( \alpha
\right) \right \vert -1\right] .  \label{37}
\end{equation}

By definition, the quantity $\delta $ is equal to zero for coherent and
squeezed vacuum states, as their WFs are non-negative. Once knowing the
Wigner function of a quantum state, we can obtain the negative part volume
of WF through numerical integration. 
\begin{figure}[tbp]
\label{Fig8} \centering \includegraphics[width=0.8\columnwidth]{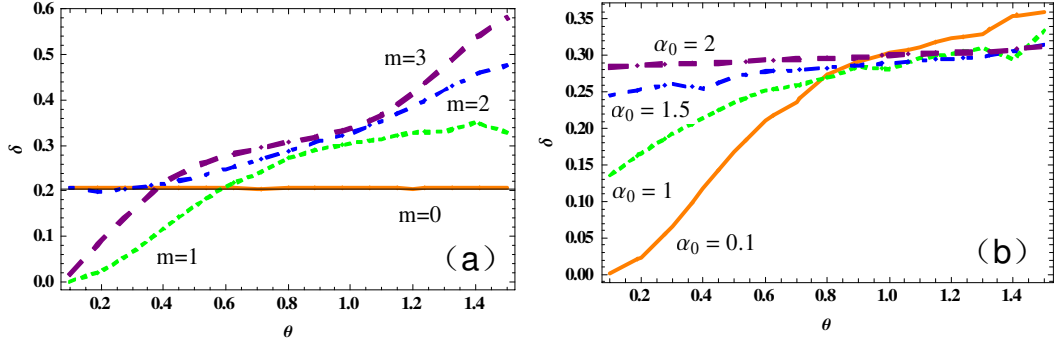}
\caption{The volume of the negative part of the WF for MCSO-OSCS as the
function of $\protect \theta $ (a) $\protect \alpha _{0}=0.1,\protect \varphi %
=0 $; (b) $m=1,\protect \varphi =0$.}
\end{figure}

In Fig. 8, we plot the negative part volume $\delta $ of WF for MCSO-OSCS as
the function of $\theta $. It is shown that the negative part volume $\delta 
$ generally increases as $\theta $ increases when $m\neq 0$. In addition, it
is interesting to note that $\delta $ is sensitive to parameter $m,$ and $%
\delta $ increases as $m$ increases when parameter $\theta $ is bigger than
a threshold (see Fig. 8(a)). In other words, the MCSO-OSCS may exhibit more
nonclassicality by increasing the value of$\ m$. Meanwhile, $\delta $
increases as the value of $\alpha _{0}$ increases when parameter $\theta $
is smaller than a threshold (see Fig. 8(b)).

\section{The decoherence of the MCSO-OSCS in a thermal environment}

When the MCSO-OSCS evolves in the thermal channel, the evolution of the
density matrix in the Born-Markov approximation and the interaction picture
can be described by the master equation \cite{47}

\begin{equation}
\frac{d\rho }{dt}=\kappa \left( \bar{n}+1\right) \left( 2a\rho a^{\dagger
}-a^{\dagger }a\rho -\rho a^{\dagger }a\right) +\kappa \bar{n}\left(
2a^{\dagger }\rho a-aa^{\dagger }\rho -\rho aa^{\dagger }\right) ,
\label{38}
\end{equation}%
where $\kappa $ represents the dissipative coefficient and $\bar{n}$ ($\bar{n%
}=\frac{1}{e^{\hbar \omega /(k_{B}T)}-1}$, $T$ is temperature.) denotes the
average thermal photon number of the environment \cite{48}. Using the
thermal entangled state representation \cite{49}, the time evolution of
distribution functions in the dissipative channels are derived \cite{50,51}.
The evolutions of the WF is governed by the following integration equation%
\begin{equation}
W\left( \gamma ,\gamma ^{\ast },t\right) =\frac{2}{\left( 2\bar{n}+1\right)
\Gamma }\int \frac{d^{2}\alpha }{\pi }W\left( \alpha ,\alpha ^{\ast
},0\right) \exp \left[ -2\frac{\left \vert \gamma -\alpha e^{-\kappa
t}\right \vert ^{2}}{\left( 2\bar{n}+1\right) \Gamma }\right] ,  \label{39}
\end{equation}%
where $\Gamma =1-e^{-2\kappa t}$ and $W\left( \alpha ,\alpha ^{\ast
},0\right) $ is the WF of the initial state. Thus the WF at any time can be
obtained by performing the integration when the initial WF is known.

Substituting Eq. (\ref{34}) into Eq. (\ref{39}), we have 
\begin{equation}
W\left( \gamma ,\gamma ^{\ast },t\right) =W_{\alpha _{0}}\left( \gamma
,\gamma ^{\ast },t\right) +W_{-\alpha _{0}}\left( \gamma ,\gamma ^{\ast
},t\right) -\left( W_{\alpha _{0}}^{\prime }\left( \gamma ,\gamma ^{\ast
},t\right) +c.c.\right) ,  \label{40}
\end{equation}%
where 
\begin{equation}
W_{\alpha _{0}}\left( \gamma ,\gamma ^{\ast },t\right) =\sum
\limits_{k=0}^{m}\sum_{l=0}^{m-k}MVU^{l}e^{-2V\left \vert \gamma -\alpha
_{0}e^{-\kappa t}\right \vert ^{2}}\left \vert H_{m-k-l}\left( -C^{\ast
}+R\alpha _{0}U+R\gamma e^{-\kappa t}V\right) \right \vert ^{2},  \label{41}
\end{equation}%
\begin{eqnarray}
W_{\alpha _{0}}^{\prime }\left( \gamma ,\gamma ^{\ast },t\right) &=&\sum
\limits_{k=0}^{m}\sum_{l=0}^{m-k}MVU^{l}e^{\left( -2\left \vert \gamma
\right \vert ^{2}V-2\left \vert \alpha _{0}\right \vert ^{2}U+2\gamma
e^{-\kappa t}\alpha _{0}^{\ast }V-2\gamma ^{\ast }e^{-\kappa t}\alpha
_{0}V\right) }  \notag \\
&&\times H_{m-k-l}\left( -B^{\ast }+R^{\ast }\alpha _{0}^{\ast }U+R^{\ast
}\gamma ^{\ast }e^{-\kappa t}V\right) H_{m-k-l}\left( B-R\alpha
_{0}U+R\gamma e^{-\kappa t}V\right) ,  \label{42}
\end{eqnarray}%
and 
\begin{eqnarray}
V &=&\frac{1}{2\bar{n}\Gamma +1},U=1-e^{-2\kappa t}V,  \notag \\
M &=&\frac{N_{m}^{-1}\left( -1\right) ^{k}2^{2l+k-m}\left( m!\right) ^{2}}{%
\pi k!l!\left( \left( m-k-l\right) !\right) ^{2}}\frac{\sin ^{k+l+m}\theta }{%
\cos ^{k+l-m}\theta }.  \label{43}
\end{eqnarray}%
Further, when $t=0,\Gamma =0$, Eq.(\ref{40}) just reduces to (\ref{34}), as
expected. 
\begin{figure}[tbp]
\label{Fig9} \centering \includegraphics[width=0.7\columnwidth]{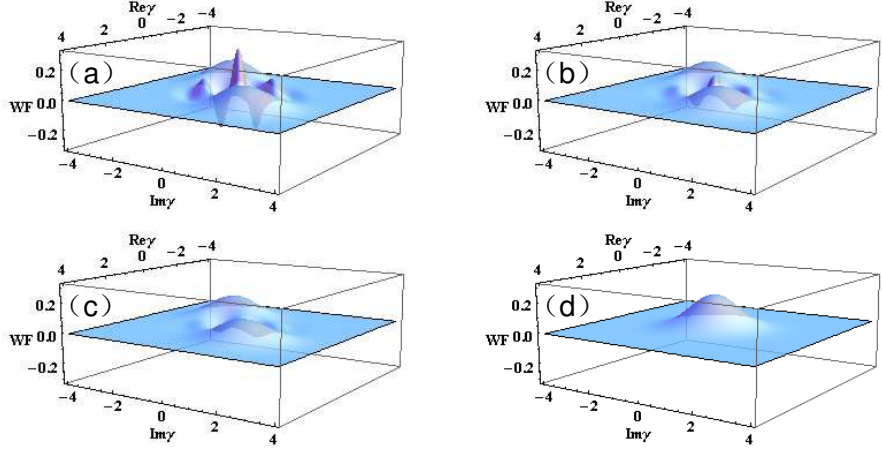}
\caption{The time evolution of Wigner function for MCSO-OSCS in the thermal
environment with $\protect \varphi =0,\protect \theta =\frac{\protect \pi }{3},%
\protect \alpha _{0}=1+i$, $m=1,\bar{n}=0.2$ (a) $\protect \kappa t=0.001$;
(b) $\protect \kappa t=0.05$; (c) $\protect \kappa t=0.1$ ; (d) $\protect%
\kappa t=3$. }
\end{figure}
\begin{figure}[tbp]
\label{Fig10} \centering \includegraphics[width=0.7\columnwidth]{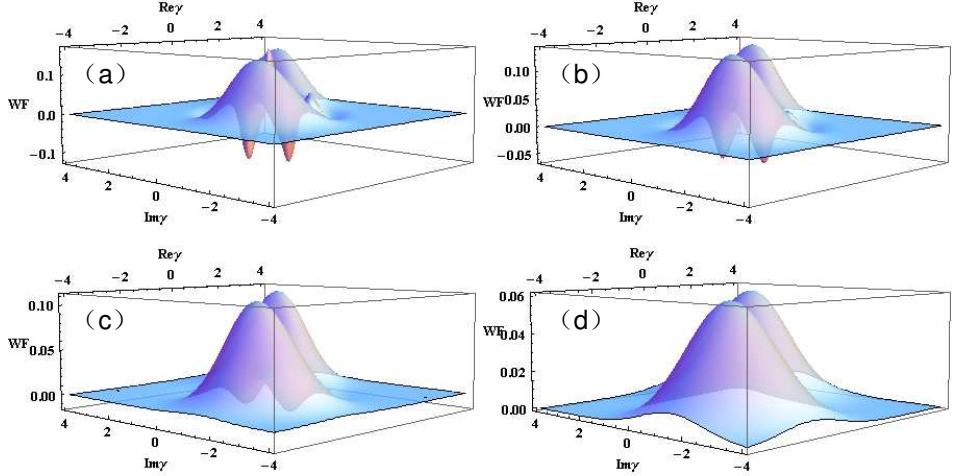}
\caption{Wigner function distributions of MCSO-OSCS in the thermal
environment for $\protect \varphi =0,\protect \theta =\frac{\protect \pi }{3},%
\protect \alpha _{0}=1+i,m=1,\protect \kappa t=0.05$ with different parameter $%
\bar{n}$ (a) $\bar{n}=0$; (b) $\bar{n}=0.5$; (c) $\bar{n}=2;$ (d) $\bar{n}=8$%
.}
\end{figure}

In order to see the decoherence of the MCSO-OSCS in the thermal
environment,\ we plot the time\ evolution of WF $W\left( \gamma ,\gamma
^{\ast },t\right) $ as a function of real and imaginary parts of $\gamma $
for some different $t$ values and for a given $m$ value (say, $m=1$) in
Figs. 9. It is shown that as time proceeds the negative part of WF and
muti-peaks vibration structure of the plot disappear gradually, and finally
the plot evolves to a wave packet structure, the figure of a Gaussian
distribution (see Fig. 9(d)), which means that\ the MCSO-OSCS has reduced to
the thermal state. In Fig. 10, we plot the picture of $W\left( \gamma
,\gamma ^{\ast },t\right) $ for some different $\bar{n}$ values and for a
given $m$ value (say, $m=1$) at the given time (say, $\kappa t=0.05$). It is
interesting to note that the negative part of WF decreases as the average
photon number $\bar{n}$ increases, i.e., the lager $\bar{n}$ the more
rapidly the nonclassicality is lost, which means that the higher the
temperature of thermal field, the more rapidly the nonclassicality of the
MCSO-OSCS is lost. This result is same as Ref. \cite{52}. \ 

\section{Conclusions}

In summary, we investigate the nonclassicality of MCSO-OSCS which is
obtained through $m$ times coherent superposition operator $a\cos \theta
+a^{\dagger }e^{i\varphi }\sin \theta $ operating on an odd-Schr\H{o}%
dinger-cat state. For arbitrary $m$ value, through IWOP technique we have
obtained an analytical expression of the normalization constant, which turns
out to be ralated with the Hermite polynomial. Then the fidelity between
MCSO-OSCS and its original OSCS is discussed. By numerical plot, it is
obvious to note that the fidelity is equal to $0$ when $m$ is odd and not
equal to $0$ when $m$ is even. The nonclassical properties of the state,
such as sub-Poissonian statistics, quadrature squeezing properties, and
photocount distribution are also discussed in details. We find that
MCSO-OSCS has more chance to exhibit sub-Poissonian statistics with bigger
value of $\theta $ in the area of $\theta \in (0,\frac{\pi }{2})$. We also
find that MCSO-OSCS can exhibit squeezing when the parameter $m$ is odd and
the angle $\theta $ is smaller, which indicates that the subtracting photon
operation is benifit to squeezing for odd $m$. Furthermore, the
nonclassicality of MCSO-OSCS is investigated in terms of WF and the negative
part volume of WF after deriving the analytical expression of WF. It is
shown that the WF of the MCSO-OSCS always has negative values which implies
the highly nonclassical properties of quantum states. The negative part
volume of WF increases as $m$ increases when $m\neq 0$ and parameter $\theta 
$ is bigger than a threshold, and increases as the value of $\alpha _{0}$
increases when parameter $\theta $ is smaller than a threshold. Especially,
the negative part volume of WF increases with the increment of parameter $%
\theta $ except the case of $m=0$.

We also investigate the decoherence of the MCSO-OSCS in terms of the
fadeaway of the negativity of WF in a thermal environment. It is shown that
nonclassicality of the MCSO-OSCS decreases as time proceeds and the
MCSO-OSCS reduces to the thermal state finally. It is also shown that
nonclassicality is influenced by the temperature of environment, the higher
the temperature is, the more rapidly the nonclassicality of the MCSO-OSCS is
lost. We wish that our results will benefit for instructing experiments, for
example, the new state would become a sufficient resource to generate a
two-mode entanglement via a beam-splitter.

\begin{acknowledgments}
This project was supported by the National Natural Science Foundation of
China (Nos.11264016, 11364022) and the Natural Science Foundation of Jiangxi
Province of China (No.20142BAB202004) as well as the Research Foundation of
the Education Department of Jiangxi Province of China (Nos.GJJ12171,
GJJ12172).
\end{acknowledgments}

\end{document}